# A Physics Model-Guided Online Bayesian Framework for Energy Management of Extended Range Electric Delivery Vehicles

Pengyue Wang, Yan Li, Shashi Shekhar and William F. Northrop*

*Abstract*—Increasing the fuel economy of hybrid electric vehicles (HEVs) and extended range electric vehicles (EREVs) through optimization-based energy management strategies (EMS) has been an active research area in transportation. However, it is difficult to apply optimization-based EMS to current in-use EREVs because insufficient knowledge is known about future trips, and because such methods are computationally expensive for large-scale deployment. As a result, most past research has been validated on standard driving cycles or on recorded high-resolution data from past real driving cycles. This paper improves an in-use rule-based EMS that is used in a delivery vehicle fleet equipped with two-way vehicle-to-cloud connectivity. A physics model-guided online Bayesian framework is described and validated on large number of in-use driving samples of EREVs used for last-mile package delivery. The framework includes: a database, a preprocessing module, a vehicle model and an online Bayesian algorithm module. It uses historical 0.2 Hz resolution trip data as input and outputs an updated parameter to the engine control logic on the vehicle to reduce fuel consumption on the next trip. The key contribution of this work is a framework that provides an immediate solution for fuel use reduction of in-use EREVs. The framework was also demonstrated on real-world EREVs delivery vehicles operating on actual routes. The results show an average of 12.8% fuel use reduction among tested vehicles for 155 real delivery trips. The presented framework is extendable to other EREV applications including passenger vehicles, transit buses, and other vocational vehicles whose trips are similar day-to-day.

## I. INTRODUCTION

Implementing an effective energy management strategy (EMS) is a key method for increasing the fuel economy of hybrid electric vehicles (HEVs) and extended range electric vehicles (EREVs). EMS's can be mainly divided into rule-based (RB) methods and optimization-based (OB) methods [1][2]. It is well known that OB methods outperform simple RB methods by a significant margin. Therefore, a preponderance of existing literature focuses on OB methods; they are introduced and discussed thoroughly in [2].

The feasibility of an EMS is mainly dependent on two factors: the computational cost and the assumptions and information needed for execution. The high fuel efficiency achieved by OB methods depends heavily on these two factors. Firstly, OB methods require considerable computational resources [2][13]. For example, in [8], X. Zeng et al. developed a stochastic model predictive control-based EMS which achieved high fuel efficiency. However, the computation time for each stochastic dynamic programming (DP) was ranged from 10 to 100 seconds, an unfeasible duration for real-world applications during vehicle operation.

Second, OB methods require either detailed trip information or broad assumptions to predict future trip information like the vehicle's second-by-second velocity profile [2]. Accurate predictions about the future are very difficult to make, especially if vehicles are controlled by human drivers in actual traffic conditions. This difficulty has motivated considerable previous research regarding driving style recognition as has been summarized in [3].

A wide variety of assumptions have been applied to incorporate future trip information to enable OB EMS methods in real-time, with varying degrees of success. In [4] and [5], future driver power demands were modeled as Markov Chains which made the real-time optimization of stochastic model predictive control (SMPC) feasible. In [6], results from DP served as training data for neural networks (NN) which were used as online controllers. The results were shown on six driving cycles which represented different driving conditions. In [7], a real-time EMS based on Pontryagin's minimum principle (PMP) was developed and tested on three standard driving cycles and one real driving cycle. X. Zeng et al. [8] modeled road grade, speed limit and stops using three transition matrices. The computational complexity of the proposed stochastic DP (SDP) algorithm was highly dependent on the sizes of the transition matrices. In [9], the real-time operation of equivalent consumption minimization strategy (ECMS) was realized through the radial basis function NN used to predict the energy demand in trip segments. Driving conditions and styles were assumed to be the same in this work. In [10], an adaptive ECMS for real-time use was developed on standard driving cycles. X. Zeng et al. [11] derived a lookup table for online using by analyzing the results from offline SDP algorithm

The information, data, or work presented herein was funded in part by the Advanced Research Projects Agency-Energy (ARPA-E) U.S. Department of Energy, under Award Number DE-AR0000795.

P. Wang, Y. Li, S. Shekhar and W. F. Northrop are with the University of Minnesota, Minneapolis, MN 55455. (email: wang6609@umn.edu; lixx4266@umn.edu; shekhar@umn.edu; wnorthro@umn.edu)

4which benefited from recognizing existing frequent routes. S. Zhang et al. [12] used a Bayes NN to predict the movement of preceding vehicles with the help of information from vehicle-to-vehicle (V2V) communication. S. Xie et al. [13] placed attention on algorithmic time efficiency in an effort to develop a low-cost controller. They compared NN and Markov chain to predict velocity and chose the latter to predict future bus trajectories which facilitated the real-time application of SMPC.

The benefits of including road grade information for ECMS and DP were shown in [14] assuming vehicle velocity was constant. S. Kermani et al. [15] assumed the route information was available in advance and showed global optimization methods can be used in real-time for a given bus route. The benefits of information provided by intelligent transportation system (ITS) were investigated in [16]. C. Sun et al. [17] used dynamic traffic feedback data and assumed that all vehicles can provide the required information. The velocity profile of the vehicle was assumed to be the same as the traffic flow, which was calculated by the average of all vehicles that on that road segment. SOC trajectories were planned by DP every 300 seconds. In [18], a real-time ECMS was enabled by a NN used for velocity prediction, using the information from V2V and vehicle-to-infrastructure (V2I). X. Qi et al. [19] assumed that a short-term velocity prediction model would be available in the future and developed an online EMS based on an evolutionary algorithm. D. Chen et al. [20] performed SOC planning for short horizon optimal control using sparse traffic information over a given route and demonstrated the near optimal fuel efficiency on standard driving cycles. Compared with DP, they reduced the computation time from hours to less than a minute. In [21], a genetic algorithm was used to reduce the computation time for optimizing power-split control parameters. In summary, to be used in actual future vehicle trips, OB EMS approaches require either more advanced transportation infrastructure (e.g., to provide dynamic traffic feedback data [17]) or to make broad assumptions about future velocity profiles (e.g., accurate short-term velocity prediction [19]).

Based on this review, it is clear that researchers have successfully demonstrated the potential of OB EMS in real-time applications using predictive methods and information from advanced transportation infrastructures. However, significant deficiencies remain to effectively use them in a human controlled transportation system, thus reducing their feasibility. With the further development of ITS, connected technology, and autonomous vehicles, OB methods may become more promising for real-world applications in the future because more accurate future trip information may be available. However, for current production vehicle configurations and transportation communication infrastructures, RB methods remain attractive for reducing vehicle fuel consumption.

RB methods are widely used industrially in a variety of applications because they are robust, easy to implement, can use simple hardware, and have low computational requirements compared to OB methods [2]. Although RB EMS's cannot achieve the highest fuel efficiency in theory, their full potential for in-use HEVs and EREVs has not been realized. Parameters in the predefined rules are usually tuned on standard driving cycles or combinations of standard driving cycles [22]. New efficient RB methods are developed on standard driving cycles [23][24]. Also, RB control strategies can be learned from results of OB methods on given driving cycles. For example, global optimal solutions on standard driving cycles were first calculated by DP, then rules were extracted from the optimal solution[25]-[27]. More discussion about the connection between RB and OB methods can be found in [2].

Results from standard driving cycles can provide valuable information about the performance and characteristics of a RB EMS. However, standard cycles cannot accurately represent the very large range of real driving vocations; for example, they have yet to be effectively applied to in-use EREVs for package delivery, the subject of this work. Parameters guiding RB EMS are set such that vehicle performance can be maintained with no impact on drivability. This results in modest improvements in fuel economy, but the risk of running out of battery in EREV vehicles is very low. Consequently, it is important to tune parameters in the predefined control rules under in-use driving cycles adaptively, considering both the risk of running out of battery and excessive fuel use by the engine. In [28], a predictive RB blended model was developed. Fuel economy improvement was demonstrated compared to a conventional RB method in a simulation environment that included uncertainty in energy demand. However, this method required distance and road type of the route in advance, which is not applicable in many real-world applications.

Data quality is also an important factor for implementing RB methods in practice. In most literature, data quality is not a consideration because standard driving cycle data [21][22] or accurate past data [12] is used. However, in real-world problems, data collected from the vehicles might have low resolution and signal noise that hinders its direct use in simulation. Preprocessing the vehicle data prior to its use in the EMS can overcome resolution and noise issues.

RB methods are compared to OB methods in the schematic given in Fig. 1, illustrating the tradeoff between feasibility for human driven vehicles and fuel efficiency improvement. A perfect model has both high feasibility and results in high fuel efficiency improvement. RB EMS methods tend to result in lower fuel efficiency improvements but are feasible to implement, whereas OB methods can attain high fuel efficiency improvement but have low feasibility due to the reasons previously discussed. For example, DP is a typical strategy employed in OB methods and can achieve a theoretical highest fuel efficiency with known detailed future trip information and at high computational expense.

This paper demonstrates an improved RB strategy for in-use EREVs by tuning RB EMS while not decreasing implementation feasibility for human driven vehicles where future velocity trajectory cannot be accurately predicted. Fig. 1 also illustrates the contribution of this work: most literature attempts to increase the feasibility of OB EMS by predicting future travel needs and assuming the availability of advanced transportation communication infrastructures. The goal of this work is to improve the fuel efficiency of a baseline vehicle RB EMS using only prior trip information.

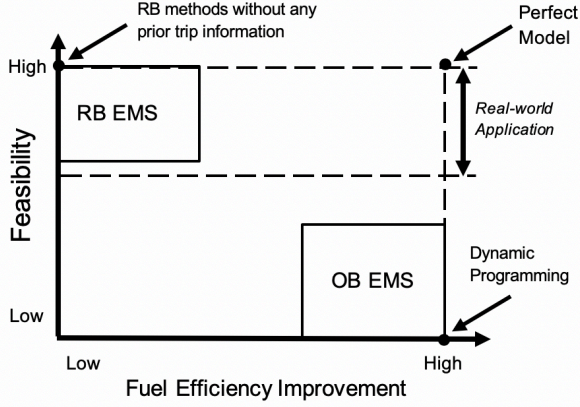

Fig. 1. Illustration comparing rule-based (RB) EMS strategies to optimization-based (OB) strategies for human driven vehicle systems.

Here, an online Bayesian algorithm is used to improve the baseline RB EMS of the vehicles. Online Bayesian algorithms are commonly used to solve sequential estimation problems with timely updates [33][35]. In RB EMS, it is important to incorporate the latest data to adaptively change parameters to avoid undesirable vehicle conditions like running out of battery energy. Online Bayesian algorithms also provide robust and stable results when a proper prior is designed, enabling conservative updates to the RB EMS even when limited historical data are available.

The outline of the paper is as follows: First, the specific EMS problem to solve is introduced in detail in section II; then, the online Bayesian algorithm is detailed in section III; section IV describes the simplified vehicle model for the EREV; the vehicle data and preprocessing module are introduced in section V; In section VI, the simplified vehicle model is first validated on real world data; designing the prior parameters for the Bayesian algorithm is then described; finally, fuel use reduction results are shown using real-world last-mile delivery fleet data.

## II. Overview of The Energy Management Strategy

The main components of the developed EMS framework are shown in Fig. 2. A database stores historical data collected from the vehicles used in the study. The preprocessing module processes available low resolution (0.2 Hz), noisy, in-use vehicle data to a higher resolution (1 Hz) with a physical checking procedure using trip distance matching. The vehicle model is used to simulate a trip. The online Bayesian algorithm is used to adaptively update the main parameter in the predefined rule for each individual vehicle using the output of the vehicle model. Computations are accomplished offline once a delivery trip was completed. The finished delivery trip data provided the information needed to update the RB EMS for the next trip. It should be noted that the word "online" in the "online Bayesian algorithm" represents the estimation of a parameter that is sequentially updated using new information [35]. It does not mean that the Bayesian algorithm runs in real-time while the vehicle is running.

The goal of the developed RB EMS is to reduce vehicle fuel consumption with the constraint that the battery SOC remains higher than 10% over the duration of the trip. For trips shorter than the all-electric-range (AER), the vehicles operate in electric-only mode. Fuel is only used by the range-extender (REx) to charge the battery when it is necessary. By approaching this goal, fossil fuel displacement and on-road emissions reduction can be achieved.

TABLE I
VEHICLE SPECIFICATIONS

|  | Parameters | Value | Unit |
|---|---|---|---|
| Vehicle | Curb weight | 5080 | kg |
|  | Typical weight | 6800 | kg |
| Motor | Maximum power | 235 | kW |
|  | Maximum torque | 2030 | Nm |
|  | Continuous power | 130 | kW |
|  | Continuous torque | 680 | Nm |
|  | Normal operating range | 0-3500 | rpm |
| Li-ion Battery | Usable capacity | 56 | kWh |
|  | Voltage range | 300-400 | V |
|  | Nominal voltage | 385 | V |
| Engine | Displacement | 0.647 | L |
|  | Compression ratio | 10.6 | / |
|  | Working power | 11 | kW |

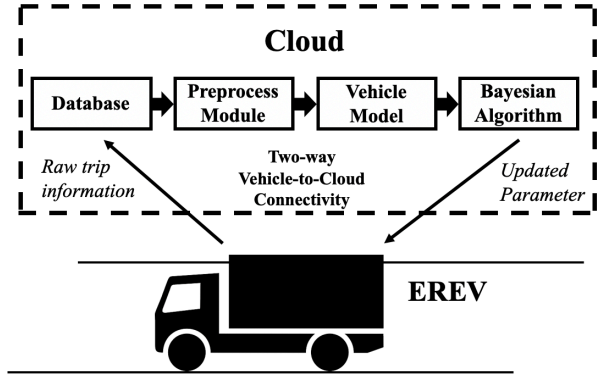

Fig. 2. Components of the physics model-guided online Bayesian framework

The configuration of the EREV powertrain studied in this work is given in Fig. 3. The motive power of the vehicle is provided by an electric motor which uses stored energy from a high-capacity battery. The internal combustion engine (ICE) serves as a REx. It is used to charge the main battery using a generator. There is no mechanical connection between the REx output shaft and the vehicle drive shaft; the ICE is completely decoupled from vehicle operation. Vehicle specifications are given in Table I.

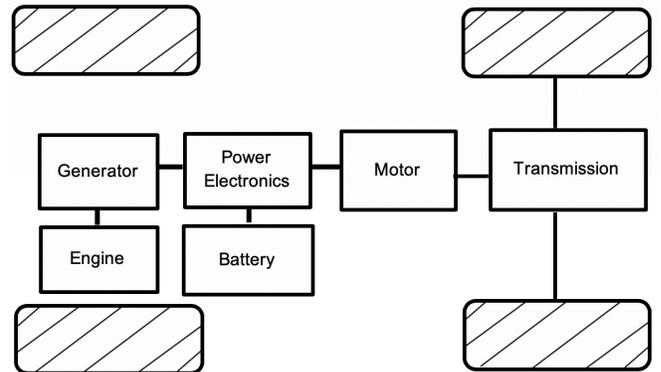

Fig. 3. Configuration of the EREV studied in this work

The RB EMS used is thermostatic, meaning that the powertrain switches between two modes to optimize fuel consumption [29]. The REx engine operates at one predetermined high efficiency speed and load condition to provide reliable, safe, and low noise operation; thus, the vehicle does not use a power-split strategy like in parallel HEV architectures [2]. The goal is to optimize when, and how frequently to operate the REx engine to minimize the fuel use and achieve a target battery state of charge (SOC) at the end of a trip, $SOC_{tev}$. Engine on-off control logic is based on SOC, which is similar to the series powertrain configuration in [30]: the engine will be turned on if the actual SOC is lower than the SOC reference value ($SOC_{ref}$). The $SOC_{ref}$ is calculated by:

$$SOC_{ref} = 100\% - (100\% - SOC_{tev}) \times \frac{e_t \times d_t}{e_e \times d_e} \quad (1)$$

In (1), $d_t$ is the distance a vehicle has traveled on a given trip and $d_e$ is the expected total trip distance for this trip. $e_t$ and $e_e$ are the energy intensity, or energy use per unit distance (kW-hr/mile) for the actual trip and the expected value, respectively. The desired $SOC_{tev}$ in this work was set 10%. The energy intensity term of the equation can be expressed through a new variable defined as $L_{set}$:

$$L_{set} = (e_e \times d_e)/e_t \quad (2)$$

Equation (1) now simplifies to (3) which is being used in the delivery fleet:

$$SOC_{ref} = 100\% \times \left(1 - 0.9 \frac{d_t}{L_{set}}\right) \quad (3)$$

The $SOC_{ref}$ decreases linearly with $d_t$ while driving. Furthermore, to reduce fuel consumption and prevent charging the battery too many times, which will degrade its life, if the calculated $SOC_{ref}$ is larger than 60%, it is set to 60% in this work. Consequently, for short trips and the beginning of long trips where SOC never drops below 60%, the REx will not operate. $SOC_{ref}$ represents how much battery energy is expected to be left when the vehicle has traveled for $d_t$ given predetermined $L_{set}$. A similar SOC reference value is also used in [32] for blended mode control for an OB EMS. The single parameter in (3) to be optimized is $L_{set}$. At the beginning of a trip, the vehicle will first operate in charge depleting (CD) mode and will switch to a charge sustaining (CS) mode such that the actual SOC follows the reference value with the aid of the REx engine. This control strategy can be classified as a blended RB method as it is designed to make the SOC achieve the lowest value at the end of the trip set by $SOC_{tev}$.

To reduce fuel consumption on a future trip, it is evident that $L_{set}$ should be preprogrammed according to trip distance and energy intensity. Ideally, if the trip distance and energy intensity of the trip is known in advance, the $L_{set}$ can be preprogrammed according to these two values so that the vehicle finishes the trip with a 10% SOC value and is charged at the depot using grid electricity at night, minimizing fuel consumption. However, $L_{set}$ is difficult to determine for at least two reasons. First, it is difficult to estimate the trip distance accurately *a priori*. Vehicles in different delivery areas have very different distributions of trip distances day-to-day. Also, for an individual vehicle, the trip distances in actual routes vary from the scheduled distance and differ day-to-day based on delivery demand, even though the vehicles might traverse the same region each day. Second, it is difficult to estimate energy intensity before a given trip as it relates to many factors like traffic condition, weather and the behavior of the driver. In application, the $L_{set}$ is generally set at a high value so that the probability of running out of battery is very small as shown in Fig. 4. Consequently, the end of route SOC is usually high which leads to excessive REx fuel consumption. For the trip example shown in Fig. 4, the REx engine would not actually be needed as the battery energy is enough for this particular trip.

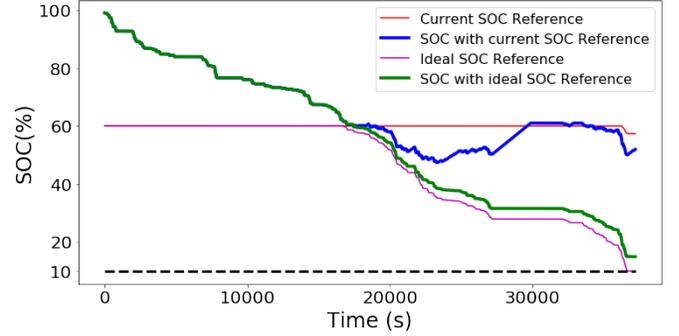

Fig. 4. Comparation of typical SOC reference and SOC value under current and ideal condition

From (3), it can be observed that if $L_{set}$ is programmed to be the actual distance (assuming that it is available before the trip), the SOC at the end of the trip will have a value of 10%. However, it can be shown using past vehicle data that even if the trip distance is known and programmed by the delivery fleet operator, it is not the best value to use for $L_{set}$ due to high energy intensity driving conditions. As the power supplied from the ICE is relatively small compared to the tractive power in a EREV, it is possible that the real SOC cannot follow the $SOC_{ref}$ when the energy intensity is high due to aggressive driving or if the vehicle has a high mass. In these cases, a very low SOC will result during the trip. As can be seen in Fig. 5, as the vehicle velocity is high at the last part of the trip, the SOC drops to a nearly 0% value. Although the final SOC is about 10%, this condition should be avoided by setting a higher $L_{set}$.

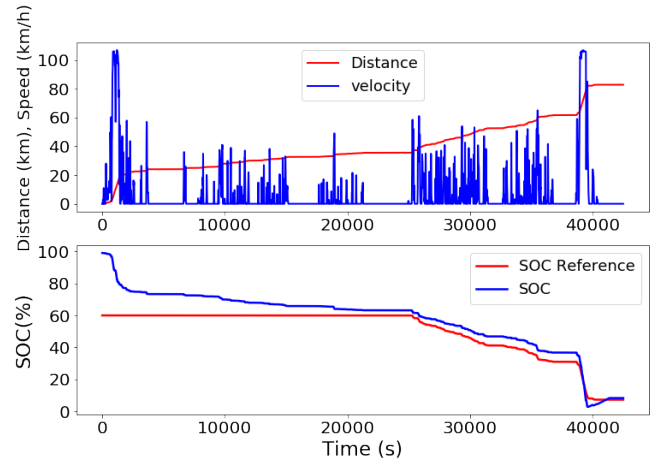

Fig. 5. A trip that the $L_{set}$ should be higher than the real distance

To find the best $L_{set}$ of a recorded trip that avoids battery SOC lower than 10% during the route, a simplified vehicle model was created. The model was run iteratively over the preprocessed velocity profile from a previous trip and outputted the minimal SOC of the trip. The value of best $L_{set}$ is the value when the minimal SOC during the trip reaches 10% (±2%). This value was then saved as the best $L_{set}$ for that trip. The vehicle model will be explained more thoroughly in section IV of this paper.

### III. Bayesian Algorithm

To program $L_{set}$ for individual vehicles, the best $L_{set}$ of all historical trips associated with one delivery vehicle can be calculated by running the simplified vehicle model. However, it is not straightforward to determine how to estimate a conservative $L_{set}$ for the next trip according all historical best $L_{set}$ as there is uncertainty about future trips. To address this uncertainty, the distribution of best $L_{set}$ of each individual vehicle using a Gaussian distribution is modeled using the cumulative density function (CDF) to get a conservative estimate of $L_{set}$ for the next trip. As shown in Fig. 6, the $L_{set}$ for next trip is the value where the CDF of the distribution equals to 0.99, which gives a very conservative prediction. For each vehicle, there is a Gaussian distribution associated with it.

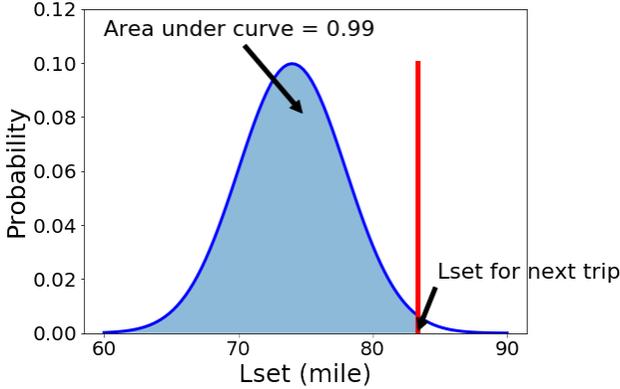

Fig. 6. Illustration of how to determine the $L_{set}$ from a distribution

However, for new vehicles or for vehicles driving new route profiles, the number of trips is very small or zero so that it is difficult to have a good estimation of the distribution and the statistical strength of such a prediction will be low. To deal with this problem, the parameters in the Gaussian distribution are estimated using a Bayesian algorithm [31]. The parameters are determined by both data and prior knowledge. Every time new trip data is available, distribution parameters are updated adaptively. Once the parameters are updated, the $L_{set}$ can be calculated conservatively by the CDF of the posterior predictive distribution.

The actual best $L_{set}$ of each vehicle is assumed to follow a Gaussian distribution with unknown mean and unknown precision: $p(L_{set}) \sim N(\mu, \lambda)$ where $\mu$ is the unknown mean and $\lambda$ is the unknown precision defined as the reciprocal of variance, $\lambda = \frac{1}{\sigma^2}$.

To simplify the notation, $L_{set}^{[N]}$ and $\hat{L}_{set}^{[N]}$ represent actual best $L_{set}$ and predicted $L_{set}$ for the $N$th trip. The following derivation mainly follows [33] and more detailed information can be found in [34] and [35].

Given historical data from $N$ trips, the likelihood can be written in the form given in (4) assuming the data are independent and identically distributed.

$$p\left(L_{set}^{[1]}, L_{set}^{[2]} \ldots L_{set}^{[N]} \middle| \mu, \lambda\right) = \frac{\lambda^{\frac{N}{2}}}{(2\pi)^{\frac{N}{2}}} \exp\left[-\frac{\lambda}{2}\sum_{i=1}^{N}\left(L_{set}^{[i]} - \mu\right)^2\right] \quad (4)$$

If $\hat{L}_{set}$ is calculated using the distribution estimated only on the historical data by maximizing the likelihood (4) yielding point estimates of $\mu$ and $\lambda$, when the size of data is small or there is no data, the calculated $\hat{L}_{set}$ will be highly unstable, leading to potentially undesirable performance of the vehicle. To solve this problem, a prior is introduced by considering the distribution of $\mu$ and $\lambda$ to make the model more conservative.

With the introduced prior distribution of $\mu$ and $\lambda$: $p(\mu, \lambda)$, the posterior distribution of $\mu$ and $\lambda$: $p\left(\mu, \lambda \middle| L_{set}^{[1]}, L_{set}^{[2]} \ldots L_{set}^{[N]}\right)$ can be determined by incorporating the historical data. As posterior $\propto$ prior $\times$ likelihood [35], the form of posterior is given by (5), which models the distribution of $\mu$ and $\lambda$ instead of just providing point estimates, i.e., instead of estimating $\mu$ and $\lambda$ as single values, the distribution of them are modeled.

$$p\left(\mu, \lambda \middle| L_{set}^{[1]}, L_{set}^{[2]} \ldots L_{set}^{[N]}\right) \propto p(\mu, \lambda) \cdot p\left(L_{set}^{[1]}, L_{set}^{[2]} \ldots \ldots L_{set}^{[N]} \middle| \mu, \lambda\right) \quad (5)$$

By introducing a prior distribution, $\mu$ and $\lambda$ is estimated based on both the information from data and prior knowledge. This can give us a more conservative estimation for small $N$. The concept of conjugate prior from Bayesian probability theory is used, which considerably simplifies the analysis. If a prior distribution is conjugate to the likelihood function of a given distribution, the posterior distribution will have the same form of distribution as the prior [35]. The conjugate prior for a Gaussian distribution with unknown mean and unknown precision is the Normal-Gamma distribution [33]: $p(\mu, \lambda) \sim Normal - Gamma(\mu_0, \kappa_0, a_0, b_0)$. There are 4 parameters in the prior. $\mu_0$ is the prior estimate of the mean. $\kappa_0$ is the size of pseudo samples from which $\mu_0$ is estimated. $a_0$ is half the size of pseudo samples from which the prior precision is estimated. $b_0$ represents the prior estimate of the precision.

With use of the conjugate prior, the posterior distribution is also Normal-Gamma: $p\left(\mu, \lambda | L_{set}^{[1]}, L_{set}^{[2]} \ldots L_{set}^{[N]}\right) \sim Normal - Gamma(\mu_N, \kappa_N, a_N, b_N)$,
where:

$$\mu_N = \frac{\kappa_0 \mu_0 + Nm}{\kappa_N}$$
$$\kappa_N = \kappa_0 + N$$
$$a_N = \frac{1}{2}(2a_0 + N)$$
$$b_N = b_0 + \frac{N}{2}s^2 + \frac{\kappa_0 N}{2\kappa_N}(m - \mu_0)^2 \quad (6)$$

$m$ and $s^2$ are the sample mean and variance of best $L_{set}$ of available trips. As can be observed in (6), the posterior mean $\mu_N$ is the weighted sum of the prior estimate of the mean and the sample mean of the available data. The posterior precision is also determined by the prior estimate and the statistic of the data. The estimate of the posterior mean and precision are based on $\kappa_N$ and $2a_N$ samples respectively, which are both the sum of number of real data and number of pseudo samples.

As the posterior distribution of $\mu$ and $\lambda$ are obtained, the next $\hat{L}_{set}$ can be predicted by using $p(\hat{L}_{set}|\mu,\lambda)$, considering all possible values of $\mu$ and $\lambda$ according to the posterior distribution $p(\mu,\lambda|L_{set}^{[1]},L_{set}^{[2]}...L_{set}^{[N]})$ by integrating over $\mu$ and $\lambda$:

$$p(\hat{L}_{set}|L_{set}^{[1]},L_{set}^{[2]}...L_{set}^{[N]})$$
$$= \iint p(\hat{L}_{set}|\mu,\lambda)p(\mu,\lambda|L_{set}^{[1]},L_{set}^{[2]}...L_{set}^{[N]})d\mu d\lambda$$
$$\sim t_{2a_N}\left(\mu_N, \frac{b_N(\kappa_N+1)}{a_N \kappa_N}\right) \quad (7)$$

Given the prior and historical data, the posterior predictive model for the next $\hat{L}_{set}$ is a t-distribution after integration. To be clear, $L_{set}$ is assumed to follow a Gaussian distribution. However, the $\mu$ and $\lambda$ are unknown such that all possibilities of $\mu$ and $\lambda$ are considered by the posterior distribution $p(\mu,\lambda|L_{set}^{[1]},L_{set}^{[2]}...L_{set}^{[N]})$ and integrated over $\mu$ and $\lambda$. The t-distribution is like an infinite sum of Gaussians [34].

Robustness is one of the main characteristics of t-distribution. It has longer 'tails' than Gaussian distribution, which means the position and shape of the t-distribution is less sensitive to outliers [35]. This property is an advantageous in this application as it can prevent the shape and position of the distribution from being influenced largely by some very short trips.

After the parameters in the prior are determined, the posterior predictive model $p(\hat{L}_{set}|L_{set}^{[1]},L_{set}^{[2]}......L_{set}^{[N]})$ is determined. This t-distribution is used to calculate the $\hat{L}_{set}$ conservatively for the next trip. After the data of the next trip is observed, the previous posterior becomes the prior and the distribution is updated according to the new observation.

The procedure for calculating $\hat{L}_{set}$ and updating parameters is described as follows:

*1) Initialization step.*

The initial t-distribution is $t_{2a_0}(\mu_0, \frac{b_0(\kappa_0+1)}{a_0\kappa_0})$, which is completely determined by the prior as there is no available trip information ($N = 0$), leading to:

$$\mu_N = \frac{\kappa_0 \mu_0 + Nm}{\kappa_N} = \mu_0$$
$$\kappa_N = \kappa_0 + N = \kappa_0$$
$$a_N = \frac{1}{2}(2a_0 + N) = a_0$$
$$b_N = b_0 + \frac{N}{2}s^2 + \frac{\kappa_0 N}{2\kappa_N}(m-\mu_0)^2 = b_0 \quad (8)$$

*2) Prediction step*

The prediction step is based on the CDF of the t-distribution as shown in Fig. 6. The value of the CDF evaluated at $\hat{L}_{set}$, is the probability that the next actual best $L_{set}$ will take a value less than or equal to the predicted $\hat{L}_{set}$:

$$CDF_{L_{set}}(\hat{L}_{set}) = P(L_{set} \leq \hat{L}_{set}) \quad (9)$$

$\hat{L}_{set}$ is determined by setting the CDF = 0.99, which means $L_{set}$ will be smaller than $\hat{L}_{set}$ with a probability of 0.99 under the assumption. From this point, it can be seen that, the calculated $\hat{L}_{set}$ will be higher than the actual ideal $L_{set}$ by a margin in most trips. For real-world driving, low $\hat{L}_{set}$ leading to a very low SOC during a trip should be avoided to a high confidence level even at the expense of smaller improvement in fuel economy.

*3) Update step*

After a new trip is observed, the parameters in the prior are updated by the parameters in the posterior; i.e., after new data is recorded, the previous posterior information becomes prior for the new information:

$$\mu_0^{new} = \mu_N^{old}$$
$$\kappa_0^{new} = \kappa_N^{old}$$
$$a_0^{new} = a_N^{old}$$
$$b_0^{new} = b_N^{old} \quad (10)$$

The parameters in the posterior are then updated according to the new data $L_{set}$ and the updated prior:

$$\kappa_N^{new} = \kappa_0^{new} + 1$$
$$\mu_N^{new} = \frac{\kappa_0^{new}\mu_0^{new} + L_{set}}{\kappa_N^{new}}$$
$$a_N^{new} = \frac{1}{2}(2a_0^{new} + 1)$$
$$b_N^{new} = b_0^{new} + \frac{\kappa_0^{new}}{2\kappa_N^{new}}(L_{set} - \mu_0^{new})^2 \quad (11)$$

After updating the parameters in the t-distribution, $\hat{L}_{set}$ for the next trip can be calculated by the prediction step.

The prior design is the most important part in the Bayesian algorithm and is specific to this RB EMS application discussed and developed in part B of section VI of this paper.

IV. SIMPLIFIED VEHICLE MODEL OF THE EREV

*A. Vehicle Dynamics*

A vehicle model is necessary as it is used to calculate the best $L_{set}$ for each trip and guides updates to the Bayesian algorithm. Also, fuel efficiency improvement is estimated by calculating the fuel use for different $L_{set}$ values. A simplified vehicle model is developed in this section for this purpose. Table II provides the parameters necessary for the vehicle model.

The vehicle force demand can be written in the following form[12]:

$$F_d = ma + c_{rr}mg\cos(\theta) + \frac{1}{2}c_d A\rho v^2 + mg\sin(\theta) \quad (12)$$

Neglecting the road grade and power estimated as $P = Fv$ gives:

$$P_d = mav + c_{rr}mgv + \frac{1}{2}c_d A\rho v^3 \quad (13)$$

The power in the case of an EREV is provided solely by the motor, which uses energy from the battery and the engine.

$$P_d = P_b \eta_{btw} + P_e \eta_{etw} \quad (14)$$

The power of battery is:

$$P_b = \frac{\left(c_{rr}mgv + \frac{1}{2}c_d A\rho v^3 + mav\right)}{\eta_{btw}} - \frac{P_e \eta_{etw}}{\eta_{btw}} \quad (15)$$

By assuming $\eta_{btw}, \eta_{etw}, m, g, c_{rr}, A, c_w, \rho$ are all constants and the fact $P_e$ is a constant (neglecting the transition process from on to off and off to on), this equation can be rewritten including the dependence on time $t$ as:

$$P_b(t) = Av(t) + Bv^3(t) + Ca(t)v(t) - D \quad (16)$$

where $A, B, C$ and $D$ are combinations of constants.

### B. Battery model

A simplified battery model is used to model the battery pack [13]:

$$P_b(t) = V_{oc}(s)I(t) - R_0(s)I^2(t) \quad (17)$$

$V_{oc}(s)$ and $R_0(s)$ depends on SOC which is denoted as $s$. The derivative of $s$ is proportional to current at the battery terminals:

$$\dot{s}(t) = -\frac{1}{Q}I(t) \quad (18)$$

If the the current is found from the battery power equation and substituted into the above equation, (19) results:

$$\dot{s}(t) = -\frac{V_{oc}(s) - \sqrt{V_{oc}^2(s) - 4R_0(s)P_b(t)}}{2R_0(s)Q} \quad (19)$$

$V_{oc}(s)$ can be modeled as a piecewise linear function of the SOC and $R_0(s)$ can be modeled as a constant $R_0(s) = R_0$. By combining (16) and (19), velocity profile can be used as an input to calculate the SOC profile step by step given the initial SOC:

$$s(t + \Delta t) = s(t) + \dot{s}(t)\Delta t \quad (20)$$

If the vehicle is stopped and the engine is on, the SOC update is simply:

$$s(t + \Delta t) = s(t) + C_c \Delta t \quad (21)$$

In this case, $\Delta t = 1s$, which means the step size is 1 second.

### C. Engine model

As the engine only works at a fixed condition, the fuel rate and engine charging power are both constant. The transition processes from off to on and on to off are neglected. So, when the engine is turned on:

$$f(t + \Delta t) = f(t) + C_f \Delta t \quad (22)$$

## V. DATA DESCRIPTION AND PREPROCESSING

### A. Data recording system and database introduction

On-board diagnostics measurement data was collected from in-use EREV delivery vehicles. Measured parameters included the vehicle's movement (e.g. velocity, distance), operation condition of powertrain components (e.g. voltage, current and SOC of battery pack, speed and torque of the traction motor, on and off status of engine) and others (e.g. ambient temperature, humidity, altitude, heater condition and signal strength). 260 parameters per vehicle in total were recorded with the timestamp and the vehicle's location every five seconds when the vehicle was running.

TABLE II
VEHICLE MODEL PARAMETERS

| Symbol | Parameter | Unit |
|---|---|---|
| $F_d$ | Total force demand of the vehicle | N |
| $c_{rr}$ | Coefficient of rolling resistance | / |
| $c_d$ | Coefficient of air resistance | / |
| $P_d$ | Total power demand of the vehicle | W |
| $P_b$ | Battery power | W |
| $P_e$ | Engine power | W |
| $\eta_{btw}$ | Efficiency from battery to wheel | / |
| $\eta_{etw}$ | Efficiency from engine to wheel | / |
| $\rho$ | Air density | kg/$m^3$ |
| $A$ | Frontal area | $m^2$ |
| $m$ | Total vehicle mass | kg |
| $g$ | Gravity constant | N/kg |
| $a$ | Acceleration | m/$s^2$ |
| $v$ | Velocity | m/s |
| $t$ | Time | s |
| $\theta$ | Road slope | rad |
| $V_{oc}$ | Open circuit voltage | V |
| $I$ | Current | A |
| $R_0$ | Battery internal resistance | ohm |
| $Q$ | Battery capacity | Ah |
| $f$ | Cumulated fuel use | L |
| $C_c$ | Battery charging rate | %/s |
| $C_f$ | Fuel consumption rate | L/s |

Data from the vehicles were stored in a secure Oracle spatial database instance with support for geometry objects and spatial indexes. The database schema consists of three main tables: Vehicle, TripSummary and DriveTrip. The Vehicle table recorded properties of each vehicle, such as the make, model and year. Every record (row) in the TripSummary table is a summary of a single delivery trip, which contains attributes such as the date, time, duration, distance and fuel use. Each record in this table is associated with a DriveTrip table. The DriveTrip table records all data of the vehicle during one delivery trip. Each row in the table describes 260 parameters of a vehicle at one spatial location with a timestamp.

Data from eight sample delivery trips by one vehicle are available in a publicly accessible data repository [36]. Each trip file contains seven columns with the physical meaning and units of each variable indicated in the first row. The seven columns contain all the information necessary for the EMS developed in this work.

### B. Data Preprocessing

Data quality is crucial to the accuracy of the simulation. However, the raw data used in this work have three challenges; first, the resolution is low. The data are recorded every 5 s. Second, there are latency in the distance and velocity data occasionally due to signal strength. Third, there are missing values. The low-resolution problem makes the data piecewise constant which is not realistic as the velocity profile should be smooth in reality. The latency problem will cause a stepped profile shape for both the velocity profile and distance profile. In addition, for the velocity profile, sometimes its value stays at zero while the distance data is increasing. In this condition, the zero-velocity value should be corrected by the distance data. To solve these problems, linear interpolation and Gaussian filter were used to increase the resolution and smooth the data. The information in the distance profile is used to correct the zero-velocity problem iteratively.

The trip-level data preprocessing procedure is as followed:
*1) Step 1*
To fill in the missing values, do zero-filling for the velocity profile and forward-filling for the distance profile;
*2) Step 2*
For both profiles, interpolate the 5 second data into 1 second data linearly;
*3) Step 3*
Use Gaussian filters to process the distance and velocity profile to get smoothed distance profile and velocity profile, the degree of smoothness is determined by $\sigma_1$ and $\sigma_2$ in the Gaussian filters; $\sigma_1$ and $\sigma_2$ are both set to be 3 to provide the degree of smoothness such that the acceleration calculated from the smoothed velocity profile and the velocity calculated from the smoothed distance profile are in normal vehicle running range.
*4) Step 4*
Calculate a new velocity profile from smoothed distance profile by second order finite difference method (for the first and last data point, velocity is zero):
$$v^{new}(t) = (d(t + \Delta t) - d(t - \Delta t))/(2\Delta t) \quad (23)$$
*5) Step 5*
Compare every point of the smoothed velocity profile and the corresponding point in the new velocity profile calculated from the smoothed distance profile and update all points that the value is 0 in the smoothed velocity profile and the value is not 0 at new velocity profile into the non-zero value multiplies by a factor ε which initializes as 1;
*6) Step 6*
Calculate new distance profile by the smoothed and corrected velocity profile and if the final distance calculated has an error smaller than 500 m, the preprocessing is finished. Otherwise, go back to step 5 and update ε according to the value of error until the stopping criteria is satisfied.

Since the actual velocity profile should be continuous and the velocity and acceleration cannot be too large, a Gaussian filter is used to infer the distance and velocity information within the 5 s data resolution. Also, the smoothing process significantly improves the data quality of the distance profile such that a new velocity profile can be found in step 4. Without the Gaussian filter, the velocity calculated from distance profile would yield unrealistic high velocities at the points where distance changes. Also, at some data points, the acceleration calculated from the unsmoothed velocity profile would be too high. Step 5 corrects for wrong velocity values. However, the velocity value calculated from the smoothed distance profile is not accurate, requiring a factor to scale the velocity. This procedure refines the data on a trip level iteratively.

Fig. 7 compares part of the raw velocity data and preprocessed data of one trip and shows the distance calculated by the raw data with zero-filling and by the preprocessed data. It can be seen that the distance calculated by the preprocessed data agrees with the raw distance data very well with a cumulated error less than 500 m. Although the smoothing process in step 3 is similar to the preprocessing in [17], physical check according to distance data is performed in the procedure, which differentiates the method from previous work.

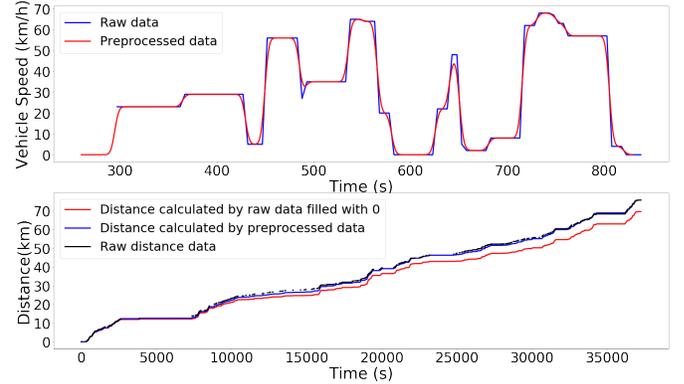

Fig. 7. Comparation of raw data and preprocessed vehicle speed and distance data

## VI. SIMULATION AND IN-USE DATA STUDY

### A. Validation of the vehicle model

The accuracy of the vehicle model is very important for the developed framework. As the engine on/off control logic is based on the SOC value, validation of the model is based on the SOC curve. Starting with the developed vehicle model, parameters were calibrated for different vehicles on each delivery area using several trips. After calibration, the model performed consistently on the other trips for the same vehicle. Errors arose from simplifications in the model including neglecting wind speed and road grade, and assuming constant vehicle component efficiencies. Also, noisy and low-resolution raw data introduced error even after the preprocessing process. Furthermore, the SOC value in the raw data itself contained some level of error as the SOC value was not measured directly. Some degree of error is inevitable in all vehicle measurement datasets. As an example, raw SOC data and simulated SOC data for one EREV are shown in Fig. 8 for four actual trips. For each simulated trip in this study, the mean relative error of the calculated SOC curve is less than 5% compared with recorded SOC curve. Considering raw data quality and model complexity as well as the goal of determining fuel consumption under different $L_{set}$, the accuracy of the model is deemed adequate.

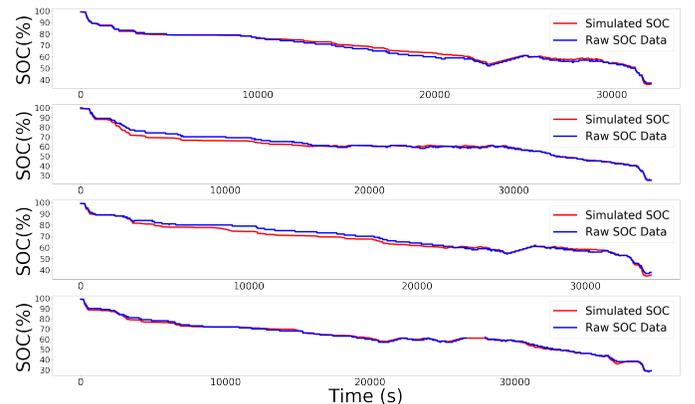

Fig.8. Validation of vehicle model by comparing the raw SOC data and simulated SOC data

## B. Designing the Prior

In this section, a prior is designed for use in the Bayesian algorithm by determining the parameters $\mu_0, \kappa_0, a_0, b_0$ from the collected vehicle data so that the $\hat{L}_{set}$ calculated is conservative, especially for small $N$ or when no data are available for new vehicles. The origin $\hat{L}_{set}$ can be used as an initial condition (100 miles). The parameters are determined using data from 78 vehicles with more than 12,000 accumulated trips in total. The 78 vehicles in different delivery areas have various mean distance and standard deviation as shown in Fig. 9.

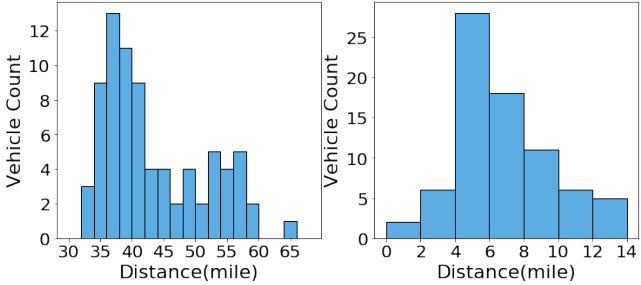

Fig.9. Distribution of the mean (left) and standard deviation (right) of trip distance for all delivery vehicles from the collected dataset.

The chosen parameters should satisfy two conditions: first, the corresponding initial distribution should yield a $\hat{L}_{set}$ that is about 100 miles; second, the series of updated $\hat{L}_{set}$ should be no lower than the actual series of $L_{set}$ for all historical trips for each vehicle. The difference between the calculated $\hat{L}_{set}$ curve and the actual best $L_{set}$ curve needs to be minimized to reduce fuel use under the constraint of $\hat{L}_{set}$ always no lower than best $L_{set}$. Considering this goal, the parameters were varied to reduce the gap between the two curves until the two curves touched for one of the vehicles. The final values of the parameters in the prior are shown in table III.

TABLE III
BAYESIAN MODEL PARAMETERS IN THE PRIOR

| Parameter | Value |
| --- | --- |
| $\mu_0$ | 74 |
| $n_{\mu_0}$ | 5 |
| $\lambda_0$ | 0.01 |
| $n_{\lambda_0}$ | 50 |

In Table III, the prior mean is 74 and is estimated from 5 pseudo samples. The prior precision is 0.01, which is estimated using 50 pseudo samples (the relation to the parameters in the prior is $\kappa_0 = n_{\mu_0}$, $a_0 = n_{\lambda_0}/2$, $b_0 = n_{\lambda_0}/2\lambda_0$).

Fig. 10 shows the calculated $\hat{L}_{set}$ curve and best $L_{set}$ curve for four vehicles, which are used to design the prior. Vehicle A and B represent about 75% of other vehicles; the calculated $\hat{L}_{set}$ is clearly higher than the actual $L_{set}$ for all trips. Vehicle C and D illustrate conditions where the calculated $\hat{L}_{set}$ and the actual $L_{set}$ are very close, which represents about 25% of all vehicles. It can be observed that there is a margin between these two curves. As the margin gets smaller, more fuel can be saved as the prediction becomes closer to the best value. However, the risk of underestimating the $L_{set}$ also increases. The designed prior minimizes the gap while avoiding the condition of calculated $\hat{L}_{set}$ being smaller than the actual $L_{set}$ for more than 12,000 trips of the 78 vehicles.

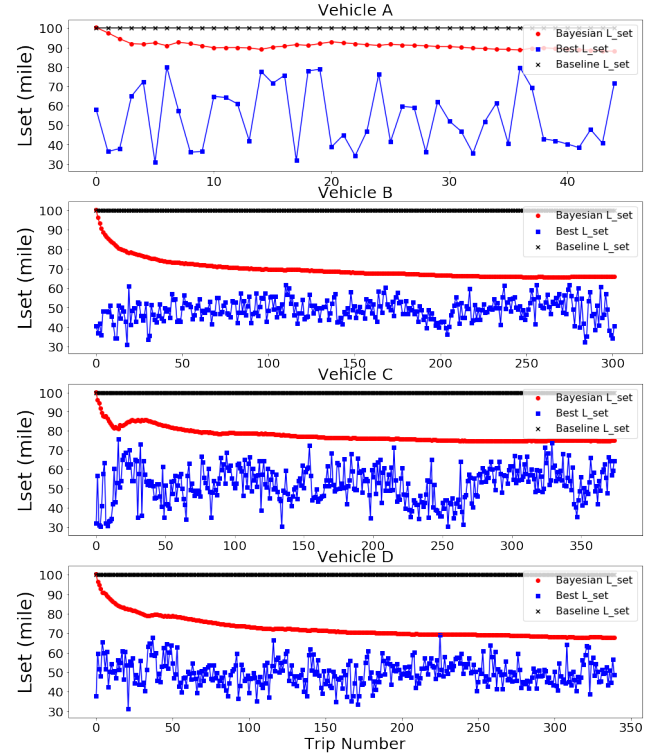

Fig. 10. Bayesian $L_{set}$ and the best $L_{set}$ curves for 4 vehicles

The initial posterior predictive distribution only determined by the prior and final distribution using all trip data of vehicle D is shown in Fig. 11. Also, the actual best $L_{set}$ data is shown in the form of histogram. It can be shown that the real best $L_{set}$ distribution can be represented by the calculated distribution well after enough data is available. Several intermediate distributions are shown in Fig. 12. It can be shown that as the number of data increases, the estimated distribution can better represent the real data distribution. When the number of data is small, the updated distribution does not significantly deviate from the prior, guaranteeing a stable and conservative $\hat{L}_{set}$. For example, even though the first best $L_{set}$ data of vehicle D is found to be less than 40 miles, which is much lower than the baseline 100 miles, the $\hat{L}_{set}$ calculated from the updated t-distribution is about 96 miles as there is only one historical trip available.

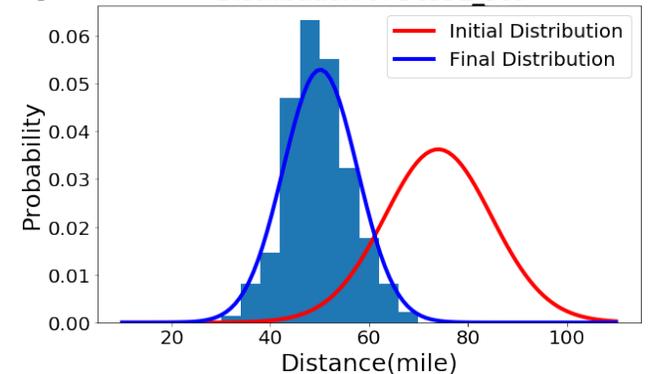

Fig. 11. Initial and final distribution of the predictive model with actual data

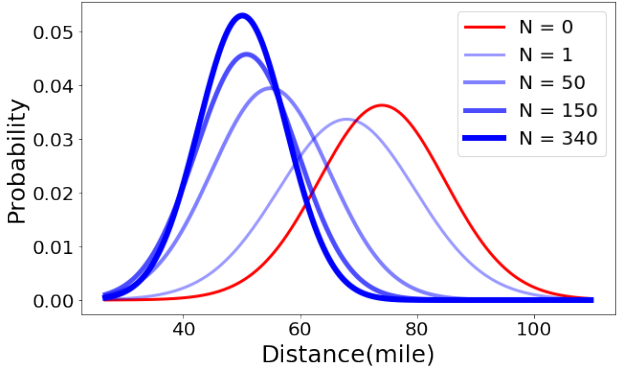

Fig. 12. Distribution estimated by different number of data

### C. Fuel use with different $L_{set}$

Fuel use for a given vehicle in a particular trip is a function of the $L_{set}$. It was observed that fuel use reduction is not guaranteed when the $L_{set}$ is lowered from the original setting. Also, the fuel use will not increase after it is higher than a particular value. For example, Fig. 13 shows the trip fuel use under different $L_{set}$ of a trip from vehicle D. It can be found that when the $L_{set}$ is higher than about 90 miles, no matter how high it is, the fuel use is the same. On the other hand, the $L_{set}$ should be lower than 90 miles to achieve some fuel reduction for this trip. The value below which fuel can be saved is different to each trip of one vehicle due to different trip distance and energy intensity.

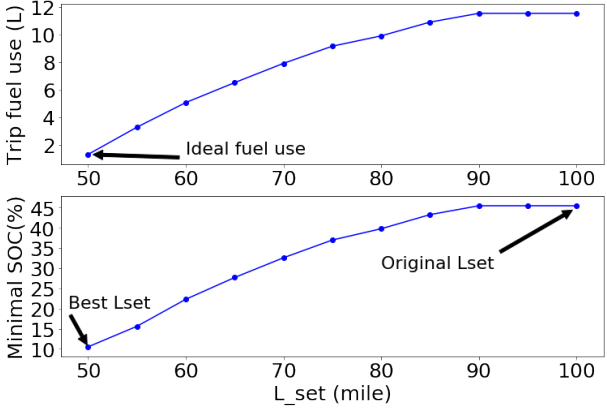

Fig. 13. Fuel use and minimal SOC under different $L_{set}$ for a trip of vehicle D

### D. Fuel efficiency improvement

The fuel efficiency improvement achieved by the EMS framework is quantified by fuel use and the mile per gallon equivalent (MPGe). MPGe is estimated by the equation [37]:

$$MPGe = \frac{distance(mile)}{fuel\ use(gallon) + \frac{electic\ energy\ use(kwh)}{33.7\left(\frac{kwh}{gallon}\right)}} \quad (24)$$

In this paper, fuel efficiency improvement is demonstrated on 13 vehicles with 155 real-world delivery trips in total. The 13 vehicles are relative new vehicles and are not used to design the prior.

In Fig. 14, the actual best $L_{set}$ and Bayesian $\hat{L}_{set}$ is shown for two of the selected demonstration vehicles. It can be observed that the update is conservative. Since the number of data points is small, the margin between Bayesian $\hat{L}_{set}$ and actual best $L_{set}$ is significant at the beginning.

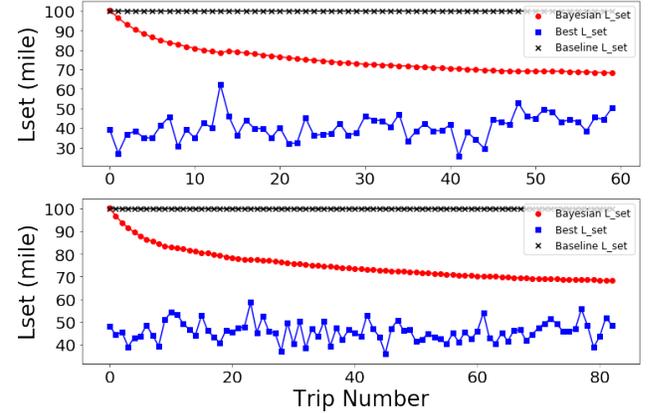

Fig. 14. Bayesian $L_{set}$, best $L_{set}$ and baseline $L_{set}$ curves for vehicle T11 (upper) and T12 (lower)

Fig. 15 illustrates how fuel is saved over the course of a real delivery trip. First, it can be seen that the simulated SOC and the actual SOC are closely aligned. Second, as the $\hat{L}_{set}$ is lowered by the developed method, the actual SOC reference value is lower than the baseline $\hat{L}_{set}$ case (100 miles). Consequently, the actual SOC follows the reference value, consuming less fuel at the last part of the trip compared with the unchanged one. Also, the minimal SOC during the trip is greater than 10%. A comparison of engine operation frequency for the last part of the same trip is shown in Fig 16. For clarity, the y-axis has two "on" positions, one for the Bayesian EMS and one for the baseline condition. It is clear from the figure that the REx engine operated much less frequently for the Bayesian EMS case, thus consuming less fuel.

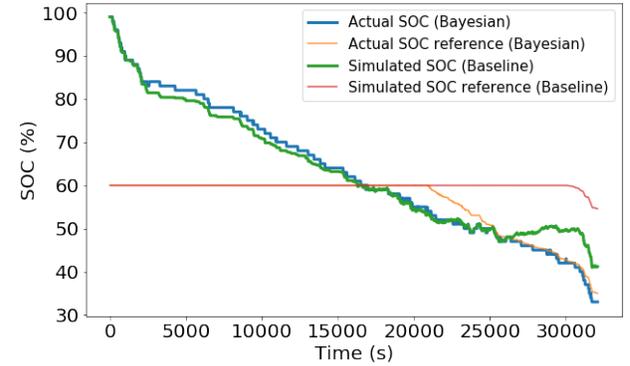

Fig. 15. Illustration of how fuel is saved in a real delivery trip

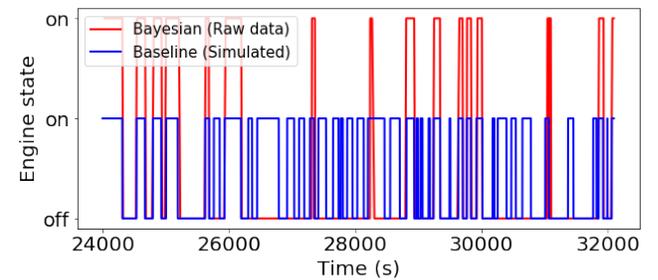

Fig. 16. Engine operation frequency for Bayesian EMS and baseline case

Fig. 17 and 18 show the detailed fuel consumption and MPGe of test delivery trips for two vehicles. It can be observed that improvement is not guaranteed when the $\hat{L}_{set}$ is lowered. The reason is that the $\hat{L}_{set}$ is not low enough as depicted in Fig. 13. All fuel reduction data for the 13 demonstration vehicles is summarized in Table IV. It can be observed that the average fuel use reduction ranges from 0% to 28.4%. The high fuel use of the original setting is due to the fact that there is no clear method to determine the $L_{set}$ for different delivery vehicles in different delivery areas. The proposed framework can effectively tune the $L_{set}$ for each vehicle in the thermostatic method by updating the estimated distribution associated with each individual vehicle every time there is a new trip data.

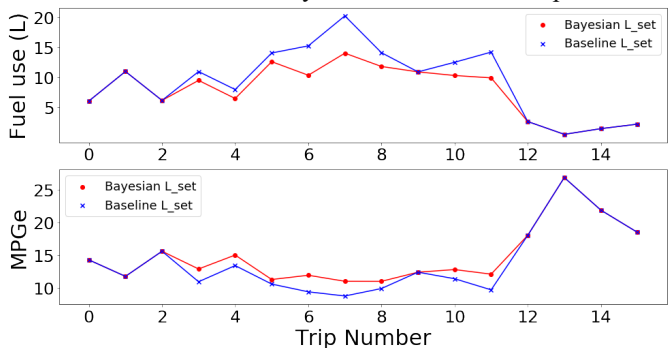
Fig. 17. Fuel use and MPGe comparison for test trips of vehicle T11

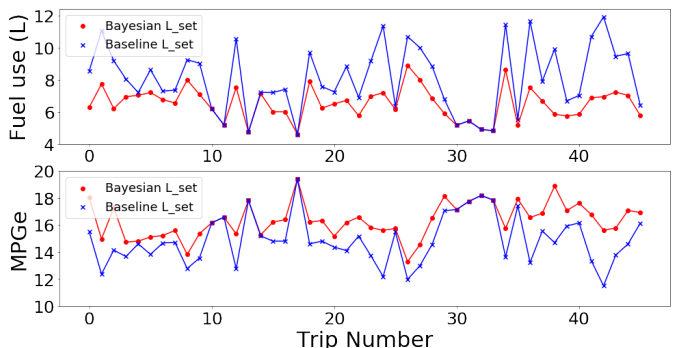
Fig. 18. Fuel use and MPGe comparison for test trips of vehicle T12

TABLE IV
Fuel efficiency improvement

| Vehicle Number | Average MPGe Improvement (%) | Average Fuel Reduction (%) | Number of Trips |
| --- | --- | --- | --- |
| T0 | 0 | 0 | 10 |
| T1 | 2.7 | 4.2 | 3 |
| T2 | 5.8 | 9.2 | 3 |
| T3 | 11.9 | 15.9 | 6 |
| T4 | 3.3 | 4.9 | 6 |
| T5 | 13.8 | 18.3 | 7 |
| T6 | 4.1 | 6.6 | 9 |
| T7 | 5.4 | 8.7 | 9 |
| T8 | 1.2 | 1.8 | 10 |
| T9 | 20.8 | 28.4 | 15 |
| T10 | 8.4 | 11.9 | 15 |
| T11 | 9.3 | 11.2 | 16 |
| T12 | 10.5 | 16.5 | 46 |
| Total | **8.9** | **12.8** | **155** |

## VII. CONCLUSION

A scalable and systematic framework to improve an in-use RB EMS including data storage, data preprocessing, vehicle model, and an online Bayesian algorithm was developed and validated using real-world driving data. Fuel use improvement was demonstrated on 13 delivery vehicles with 155 real world delivery trips in total. An average of 12.8% improvement in fuel usage is shown in practice. The developed RB EMS framework provides an immediate and feasible solution for fuel use reduction for current in-use EREV delivery vehicles and can be extended to other vehicle architectures and vocations. The developed Bayesian framework does not require high predictive capabilities or high computational expense. The RB EMS framework can also be applied to other EREV applications whose trips are similar day-to-day. As the developed algorithm programs $\hat{L}_{set}$ only at the beginning of a trip, there is potential to further reduce fuel use if it is set dynamically using real-time information during the trip; this will be the subject of future work.

ACKNOWLEDGMENT

The information, data, or work presented herein was funded in part by the Advanced Research Projects Agency-Energy (ARPA-E) U.S. Department of Energy, under Award Number DE-AR0000795. The views and opinions of authors expressed herein do not necessarily state or reflect those of the United States Government or any agency thereof.